\documentclass[journal]{IEEEtran}
\usepackage{graphics} 
\usepackage{epsfig} 
\usepackage{mathptmx} 
\usepackage{amsmath} 
\usepackage{upgreek}
\usepackage{float}

\ifCLASSINFOpdf
\else
\fi
\begin{document}
%
\title{Tunable anisotropic absorption in hyperbolic metamaterials based on black phosphorous/dielectric multilayer structures}
%
%
%

\author{Shuyuan~Xiao,~Tingting~Liu,~Le~Cheng,~Chaobiao~Zhou,~Xiaoyun~Jiang,~Zhong~Li,~Chen~Xu
\thanks{This work is supported by the National Natural Science Foundation of China (Grant No. 61775064, 11847100 and 11847132), the Fundamental Research Funds for the Central Universities (HUST: 2016YXMS024) and the Natural Science Foundation of Hubei Province (Grant No. 2015CFB398 and 2015CFB502).}
\thanks{S. Xiao is with the Institute for Advanced Study, Nanchang University, Nanchang 330031, China (email: syxiao@ncu.edu.cn).}
\thanks{T. Liu is with the Laboratory of Millimeter Wave and Terahertz Technology, School of Physics and Electronics Information, Hubei University of Education, Wuhan 430205, China (email: ttliu@hue.edu.cn).}
\thanks{L. Cheng, C. Zhou and X. Jiang are with the Wuhan National Laboratory for Optoelectronics, Huazhong University of Science and Technology, Wuhan 430074, China (email: chengle@hust.edu.cn; zcbiao586@hust.edu.cn; xyjiang@hust.edu.cn).}
\thanks{Z. Li is with the Center for Materials for Information Technology, and also with the Department of Physics and Astronomy, The University of Alabama, Tuscaloosa 35487, USA (email: zli73@crimson.ua.edu).}
\thanks{C. Xu is with the Department of Physics, New Mexico State University, Las Cruces 88001, USA (email: chenxu@nmsu.edu).}
\thanks{Manuscript received Jan 17, 2019; revised MM DD, 2019.}}

%
%

\markboth{Journal of \LaTeX\ Class Files,~Vol.~14, No.~8, August~2015}%
{Shell \MakeLowercase{\textit{et al.}}: Bare Demo of IEEEtran.cls for IEEE Journals}
%



\maketitle

\begin{abstract}
Black phosphorus (BP) as a promising two-dimensional material with extraordinary optical properties constitutes an excellent building block in multilayer hyperbolic metamaterials. In this work, we design a multilayer structure composed of BP/dielectric layer stacking unit cells patterned on a gold mirror and theoretically demonstrate the tunable anisotropic absorption in the infrared regime. The electric dipole resonance between the adjacent unit cells drives the structure in the critical coupling state, and impedance of the structure matches to that of free space, showing the perfect absorption for one polarization direction, while the impedance mismatch for the other polarization direction leads to only 8.2$\%$ absorption at the same wavelength. The anisotropic absorption response of the proposed structure can be attributed to the intrinsic anisotropy of BP, which exhibits few dependence on the incident angle. Furthermore, we investigate the tunable optical absorption of the proposed structure with the electron doping of BP and the geometric parameters. These results demonstrate great potentials of BP in constituting multilayer hyperbolic metamaterials, and open up avenues in designing anisotropic metadevices with tunable spectral and polarization selectivity in the infrared regime.
\end{abstract}


%
\IEEEpeerreviewmaketitle

\section{Introduction}\label{sec1}
%
%
%
%
\IEEEPARstart{M}{etamaterials} with artificially structured subwavelength building blocks have attracted substantial attention due to their unprecedented flexibility in light manipulation\cite{liu2011metamaterials}. Recently, the research on metamaterials has progressed to tunable and functional metadevice applications, as well as discovering novel structures with exceptional properties inaccessible to conventional metamaterials. Among the emerging varieties of metamaterials, hyperbolic metamaterials are most distinguished thanks to their extremely anisotropic feature enabling the hyperbolic isofrequency dispersion\cite{poddubny2013hyperbolic, sun2014indefinite}. Such metamaterials are good analogues of traditional optical crystals with one diagonal component of the permittivity tensors exhibiting the opposite sign to the other two diagonal components, which can be fulfilled by constricting the motion of free electrons within one or two spatial dimensions, practically, by periodically arranged metal/dielectric multilayers and metallic nanowire arrays. The permittivity tensors of hyperbolic metamaterials can be well defined by the effective medium theory, and expediently tuned by changing the filling ratios in addition to the material and other geometric parameters, providing a much more flexible control of the resonant responses for a wide range of applications in different spectral bands. Moreover, owing to the hyperbolic isofrequency dispersion, these simple and elegant structures possess unique optical and physical properties, such as supporting high propagation wave vectors and enhancing photonic density of states, and offer an excellent platform in tailoring light-matter interaction, which facilities prospective applications in high-resolution imaging and lithography\cite{jacob2006optical, lu2012hyperlenses, liang2015squeezing}, spontaneous emission enhancement\cite{lu2014enhancing, roth2017spontaneous, rustomji2017measurement}, ultrasensitive biosensing\cite{kabashin2009plasmonic, sreekanth2016extreme, sreekanth2016enhancing}, and broadband perfect absorption\cite{zhou2014experiment, yin2015ultra, chang2016metasurface}.

In the last decade, atomically thin two-dimensional (2D) materials which show remarkable electronic, optical, mechanical and thermal properties, have received great research interest\cite{xia2014two, he2016further, xiao2017strong, xiao2018active, li2018wavelength}. Graphene as one of the most popular 2D materials, which can take over the role of metal in providing an inductive layer with negative permittivity, constitutes an excellent building block for multilayer structure of hyperbolic metamaterials\cite{iorsh2013hyperbolic, chang2016realization}. The graphene-based hyperbolic metamaterials have been extensively investigated for light manipulation at nanoscale, empowering different applications in negative refraction\cite{sreekanth2013negative, barzegar2016study}, light confinement\cite{othman2013graphene, he2013broadband, su2015terahertz} and slow light\cite{jia2015tunable, tyszka2017tunable}. More recently, a newly emerging 2D material, black phosphorus (BP) which has been isolated from bulk-BP by mechanical exfoliation and plasma thinning, shows high carrier mobility, relatively low loss and flexible tunability\cite{low2014tunable, ling2015renaissance, liu2018dynamical}. Similar to graphene, BP presents metallic behavior with negative permittivity in the mid- and far-infrared regime\cite{low2014plasmons, gonccalves2017hybridized, lu2017strong, nong2018strong, qing2018tailoring}, thus offering capabilities to replace the metal in metal/dielectric multilayer hyperbolic metamaterials. Moreover, BP with a direct and layer-sensitive bandgap from 0.3 eV to 2 eV shows an attractive advantage to realize high on-off ratios\cite{lu2015bandgap, correas2016black, li2017tunable}, which cannot be achieved in the zero-bandgap graphene. More importantly, the asymmetric crystal structure of BP brings about the highly in-plane anisotropy, leading to extremely polarization-dependent electronic and optical properties that may open avenues for novel functional devices\cite{liu2016localized, xiong2017strong, song2018biaxial, yuan2018highly, hong2018towards, feng2019perfect}. However, the BP-based hyperbolic metamaterials have been rarely considered, and the fundamental properties and the potential applications on this aspect have yet to be fully investigated.

In this work, we propose a novel class of hyperbolic metamaterials based on BP/ dielectric multilayer structures for the realization of the tunable anisotropic absorption. To our knowledge, it is the first time to take BP instead of metal or graphene as the building block for multilayer hyperbolic metamaterials. Due to the intrinsic anisotropy of BP, the proposed structure shows perfect absorption for one polarization direction while the absorption of 8.2$\%$ for the other polarization direction at the same wavelength. Moreover, the absorption response of the proposed structure can be tuned by varying the electron doping of BP, the thickness and the number of the BP/ dielectric bilayer in the unit cell. This work confirms the potential of BP as an excellent building block in multilayer hyperbolic metamaterials, and demonstrates the promising application in tunable anisotropic metadevices.

\section{The geometric structure and numerical model}\label{sec2}
\begin{figure}[h]
\centering
\includegraphics
[scale=0.45]{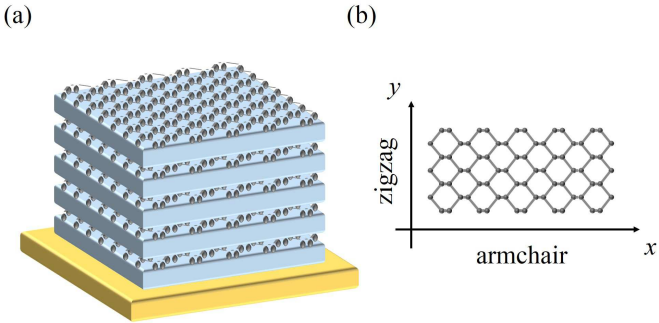}
\caption{\label{fig:1} (a) The schematic of the hyperbolic metamaterials composed of BP/dielectric layer stacking unit cells patterned on a gold mirror. (b) The top view of the lattice structure of BP.}
\end{figure}

Fig.~\ref{fig:1}(a) depicts the schematic of our proposed hyperbolic metamaterials, which are composed of BP/dielectric layer stacking unit cells patterned on a gold mirror. The period distance between unit cells is $P=500$ nm and the cubic geometry is $L=400$ nm in length of each side. The thickness of each BP/dielectric bilayer is denoted by $t=t_{BP}+t_{d}$, with thickness for BP layer and dielectric layer as  $t_{BP}=1$ nm and  $t_{d}=79$ nm, respectively. The thickness of gold mirror is set as $t_{Au}=200$ nm. The number of the BP/ dielectric bilayer in each unit cell is represented by $N=20$. The permittivity of dielectric is $\varepsilon_{d}=2$. The permittivity of Au plane is derived from Drude model with the plasmon frequency $\omega_{p}=1.37\times10^{16}$ rad/s and the collision frequency $\gamma_{p}=4.08\times10^{13}$ rad/s\cite{ordal1985optical}. In this configuration, the transmission is eliminated because the thick gold mirror prevents the propagation of incident light. 

The atoms in monolayer BP are covalently bonded to form a unique puckered honeycomb structure through $sp^{3}$ hybridization. The asymmetric crystal structure of BP, i.e. $x$ direction armchair edge and $y$ direction zigzag edge, is depicted in Fig.~\ref{fig:1}(b), leading to remarkable in-plane anisotropic electron dispersion and direction-dependent conduction. In the mid- and far-infrared regime, the surface conductivity of BP can be described using the semi-classical Drude model\cite{liu2016localized, xiong2017strong}
\begin{equation}
\label{eq:1}
    \sigma_{jj}=\frac{iD_{j}}{\pi(\omega+\frac{i\eta}{\hbar})},~j=x,y,
\end{equation}
where $j=x,y$ represents the $x$ and $y$ directions, respectively, $D_{j}$ is the Drude weight, $\omega$ is the incident light frequency, $\eta=10$ meV is the electron relaxation rate, $\hbar$ is the reduced Planck's constant. The parameter of $D_{j}$ is described by
\begin{equation}
\label{eq:2}
    D_{j}=\frac{\pi e^{2}n}{m_{j}},~j=x,y,
\end{equation}
where $e$ is the electron charge, $n$ is the electron doping, $m_{cx}=0.15m_{0}$ and $m_{cy}=0.7m_{0}$ ($m_{0}$ is the static electron mass) are the in-plane electron's effective mass along the $x$ and $y$ directions, respectively\cite{low2014tunable}. Hence, the equivalent relative permittivity of BP in three directions can be derived by
\begin{equation}
\label{eq:3}
    \varepsilon_{jj}=\varepsilon_{r}+\frac{i\sigma_{jj}}{\varepsilon_{0}\omega t_{BP}},~j=x,y,z,
\end{equation}
where $j=x,y,z$ represents the $x$, $y$ and $z$ directions, respectively. $\varepsilon_{r}=5.76$ is the relative permittivity of BP, $\varepsilon_{0}$ is the vacuum permittivity.

\begin{figure}[htbp]
\centering
\includegraphics
[scale=0.35]{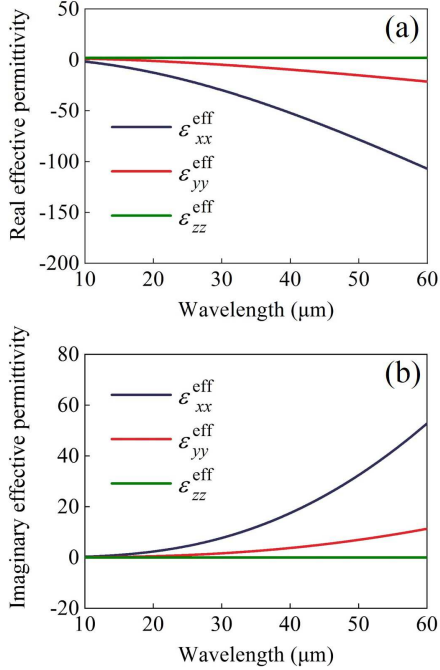}
\caption{\label{fig:2} (a) The real parts and (b) the imaginary parts of the effective permittivity tensors along $x$, $y$ and $z$ directions for the multilayer hyperbolic structure.}
\end{figure}

The subwavelength BP/dielectric multilayer structure can be treated as an effective homogeneous medium, and the effective permittivity tensors along $x$, $y$ and $z$ directions are determined by\cite{agranovich1985notes}
\begin{equation}
\label{eq:4}
    \varepsilon_{jj}^{eff}=\frac{\varepsilon_{jj}t_{BP}+\varepsilon_{d}t_{d}}{t_{BP}+t_{d}},~j=x,y,
\end{equation}
\begin{equation}
\label{eq:5}
    \varepsilon_{jj}^{eff}=\frac{\varepsilon_{jj}\varepsilon_{d}(t_{BP}+t_{d})}{t_{BP}\varepsilon_{d}+t_{d}\varepsilon_{BP}},~j=z,
\end{equation}
According to Eqs.~(\ref{eq:4})-(\ref{eq:5}), the real and imaginary parts of the three effective permittivity tensors are depicted in Fig.~\ref{fig:2}(a) and (b), respectively. In the initial setup, a moderate electron doping of $n=5\times10^{13}$ cm$^{-2}$ is considered, and we can observe $\varepsilon_{xx}^{eff},\varepsilon_{yy}^{eff}<0$ and $\varepsilon_{zz}^{eff}>0$ are satisfied in the infrared regime of interest. Therefore the hyperbolic dispersion properties are expected in the multilayer structure. By changing material properties and geometry parameters, the effective permittivity tensors can be expediently tuned, which provides an unprecedented degree of freedom to control the resonant responses compared with conventional metamaterials.

The absorption responses of the proposed hyperbolic metamaterials are investigated via simulations using finite difference time domain (FDTD) method via the commercial package FDTD Solutions, Lumerical Inc., Canada. In the calculations, the moderate mesh grid is adopted to make good tradeoff between accuracy, memory requirements and simulation time. The linearly polarized plane wave is incident along the $-z$ direction, and the periodic boundary conditions are used in the $x$ and $y$ directions and the perfect matching layer conditions are adopted in the $z$ direction.

\section{Results and discussions}\label{sec3}
\begin{figure}[htbp]
\centering
\includegraphics
[scale=0.35]{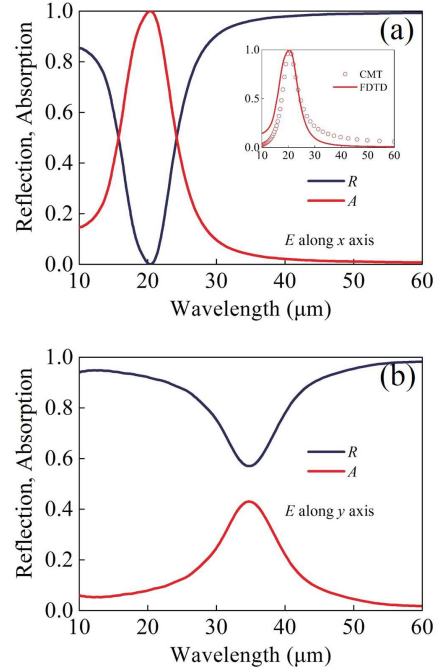}
\caption{\label{fig:3} The anisotropic absorption and reflection spectra for electric field $E$ along (a) $x$ and (b) $y$ directions under normal incidence in the proposed hyperbolic metamaterials. In the inset of (a), FDTD numerical simulation (red line) and CMT theoretical analysis (red circle) under the critical coupling condition for the electric field $E$ along $x$ direction.}
\end{figure}

Fig.~\ref{fig:3}(a) and (b) present the simulated spectra of the proposed BP/dielectric metamaterials for electric field $E$ along $x$ and $y$ directions under normal incidence. The total absorption is 100$\%$ at the resonance wavelength of 20.41 $\upmu$m for $E$ along $x$ direction, achieving the perfect absorption. On the other hand, a weak absorption of only 8.2$\%$ is observed for $E$ along $y$ direction at the same wavelength, and the strongest absorption of 43$\%$ appears at the resonance wavelength of 34.90 $\upmu$m. Therefore, the extremely anisotropic absorption can be well observed for different polarization directions.

The extremely anisotropic absorption can be interpreted with the impedance match associated with the critical coupling condition. According to the coupled mode theory (CMT), the structure can be considered as a resonator with the amplitudes of the input and output waves of $u$ and $y$ at the resonance frequency $\omega_{0}$. The mode radiative loss rate and the material loss rate are assigned as $\gamma_{e}$ and $\delta$, respectively. The reflection coefficient of the structure can be described by\cite{piper2014total, jiang2017tunable}
\begin{equation}
\label{eq:6}
    \Gamma=\frac{y}{u}=\frac{i(\omega-\omega_{0})+\delta-\gamma_{e}}{i(\omega-\omega_{0})+\delta+\gamma_{e}},
\end{equation}
and the absorption is calculated as $A=1-|\Gamma|^{2}$,
\begin{equation}
\label{eq:7}
    A=\frac{4\delta\gamma_{e}}{(\omega-\omega_{0})^{2}+(\delta+\gamma_{e})^{2}}.
\end{equation}
When the structure is driven on the resonance ($\omega=\omega_{0}$), and the mode radiative loss rate and the material loss rate are the same ($\delta=\gamma_{e}$), which fulfills the critical coupling condition, the reflection from the structure vanishes, and all the incident light can be absorbed. As shown in the inset of Fig.~\ref{fig:3}(a), the theoretical analysis (CMT) and the numerical simulation (FDTD) are in good agreement in the vicinity of the resonance for the electric field $E$ along $x$ direction. The slight deviation occurs far away from the resonance, which can be attributed to that the CMT assumes a lossless resonator off the resonance.

From the macroscopical sight, when the critical coupling condition is fulfilled, the effective impedance of the structure should be equal to that of the free space ($Z_{0}=1$). For the proposed metamaterials composed of BP/dielectric multilayer structure supported on the gold mirror, the effective impedance $Z$ can be described by\cite{smith2005electromagnetic, szabo2010unique}
\begin{equation}
\label{eq:8}
    Z=\frac{(T_{22}-T_{11})\pm\sqrt{(T_{22}-T_{11})^{2}+4T_{12}T_{21}}}{2T_{21}},
\end{equation}
where $T_{11}$, $T_{12}$, $T_{21}$ and $T_{22}$ are the element values of the transfer (T) matrix. The two solutions of the effective impedance describe the two different paths of the light propagation, for example, the plus sign corresponds to the positive direction. The elements of T matrix can be calculated from the scattering (S) matrix elements written as
\begin{equation}
\label{eq:9}
    T_{11}=\frac{(1+S_{11})(1-S_{22})+S_{21}S_{12}}{2S_{21}},
\end{equation}
\begin{equation}
\label{eq:10}
    T_{12}=\frac{(1+S_{11})(1+S_{22})-S_{21}S_{12}}{2S_{21}},
\end{equation}
\begin{equation}
\label{eq:11}
    T_{21}=\frac{(1-S_{11})(1-S_{22})-S_{21}S_{12}}{2S_{21}},
\end{equation}
\begin{equation}
\label{eq:12}
    T_{22}=\frac{(1-S_{11})(1+S_{22})+S_{21}S_{12}}{2S_{21}}.
\end{equation}
In order to achieve the perfect absorption, the effective impedance of the whole structure needs to match to that of the free space, which implies the value of $Z$ in Eq.~(\ref{eq:8}) as close to 1 as possible. The effective impedance of the structure is equal to the free space impedance at the resonance wavelength of 20.41 $\upmu$m for $E$ along $x$ direction, i.e. $Z=0.99-0.02i$, while the effective impedance is calculated as $3.49\times10^{-4}-0.16i$ at the same wavelength and as $0.14-0.04i$ at the resonance wavelength of 34.90 $\upmu$m for $E$ along $y$ direction. Due to the anisotropic property of BP along $x$ and $y$ directions, the impedance match associated with the critical coupling condition can be realized in only one direction at a time, giving rise to the anisotropic absorption as shown in Fig.~\ref{fig:3}.

\begin{figure}[htbp]
\centering
\includegraphics
[scale=0.45]{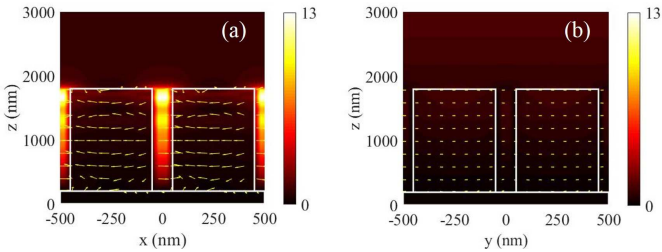}
\caption{\label{fig:4} The electric field distributions at the wavelength of 20.41 $\upmu$m for $E$ along (a) $x$ and (b) $y$ directions.}
\end{figure}

To further clarify the physical mechanism of the anisotropic absorption behaviors in the proposed structure, we simulate the electric field distributions at the wavelength of 20.41 $\upmu$m for $E$ along $x$ and $y$ directions. As shown in Fig.~\ref{fig:4}(a), the pronounced confinement of electric field between the adjacent unit cells can be observed for the electric field $E$ along $x$ direction. The divergence and convergence of the electric field exhibit a clear tendency to follow the polarization of the incident plane wave, and the positive and negative charges accumulate at the left and right sides of  the gaps, which are typical characteristics of the electric dipole resonance in the hyperbolic metamaterials. When the structure is driven on the resonance, the mode radiative loss rate and the material loss rate are the same, which fulfills the critical coupling condition. The effective impedance of the proposed structure shows a great match to that of the free space, leading to the perfect absorption of the incident plane wave for $E$ along $x$ direction. In Fig.~\ref{fig:4}(b), the weak confinement of electric field can be observed at the same wavelength for $E$ along $y$ direction, the structure is not in the critical couling state, corresponding to the weak absorption. According to the above analysis, it can be concluded that the critical coupling or not directly causes the impedance match or mismatch of the whole structure to that of free space, and leads to the anisotropic absorption, which is rooted in the asymmetric crystal structure of BP.

\begin{figure}[htbp]
\centering
\includegraphics
[scale=0.45]{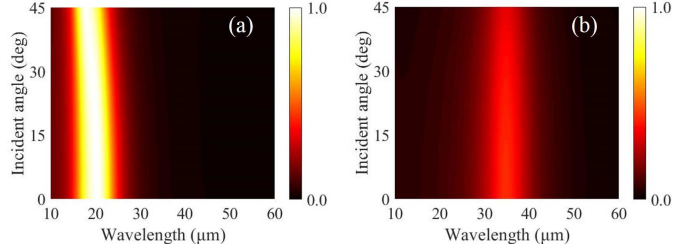}
\caption{\label{fig:5} The angular dependence of the absorption spectra for $E$ along (a) $x$ and (b) $y$ directions.}
\end{figure}

Fig.~\ref{fig:5}(a) and (b) present the dependence of the absorption spectra on the incident angle for $E$ along $x$ and $y$ directions, respectively. For $E$ along $x$ direction, the near perfect absorption can be observed within a large incident angle from 0 to 45 degree. The resonance wavelength shows a slight blue shift as the increase of the incident angle, which may be induced by the unsymmetrical electric field component of the oblique incident plane wave. For $E$ along $y$ direction, the absorption efficiency gradually decreases when the incident angle increases, and the absorption decreases rapidly when the incident angle exceeds 30 degree. Hence, the proposed BP/dielectric multilayer structure can well maintain the anisotropic absorption behaviors at the incident angle from 0 to 45 degree for $E$ along $x$ and $y$ directions.

By changing material properties and geometry parameters, we can get flexible control of the absorption responses of the proposed hyperbolic metamaterials. In the following, we investigate the tunable anisotropic absorption of the structure by varying the electron doping $n$ of BP layer, the thickness $t$ and the layer number $N$ of the BP/dielectric bilayer.

\begin{figure}[htbp]
\centering
\includegraphics
[scale=0.35]{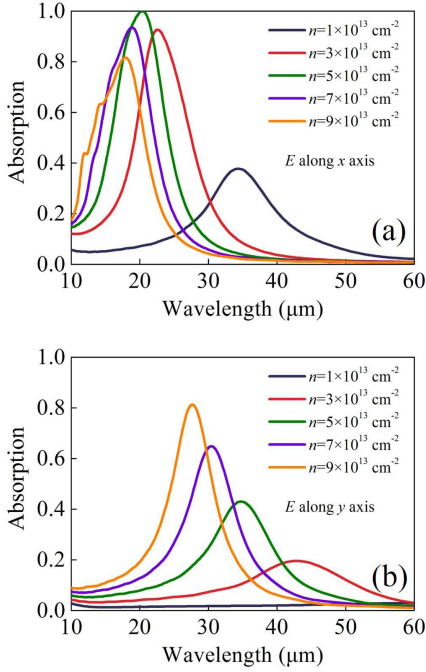}
\caption{\label{fig:6} The anisotropic absorption spectra for electric field $E$ along (a) $x$ and (b) $y$ directions under normal incidence with various electron doping of BP ($n$ ranging from $1\times10^{13}$ cm$^{-2}$ to $9\times10^{13}$ cm$^{-2}$) in the proposed hyperbolic metamaterials.}
\end{figure}

According to Eqs.~(\ref{eq:1})-(\ref{eq:5}), the electron doping $n$ directly determines the surface conductivity of BP, then influences the effective permittivity tensors and the absorption behavior of the whole structure. In Fig.~\ref{fig:6}(a) and (b), the absorption spectra are plotted at different electron doping $n$ ranging from $1\times10^{13}$ cm$^{-2}$ to $9\times10^{13}$ cm$^{-2}$, where the thickness $t=80$ nm and the number $N=20$ of the BP/dielectric bilayer are in line with the initial setup. The change in the electron doping $n$ of BP causes obvious variations in the absorption response. For $E$ along $x$ direction, the absorption peak of the proposed structure gradually increases from 37.78$\%$ to 100$\%$ as $n$ increases from $1\times10^{13}$ cm$^{-2}$ to $5\times10^{13}$ cm$^{-2}$, and then decreases to 81.58$\%$ with the further increase of $n$ to $9\times10^{13}$ cm$^{-2}$. For $E$ along $y$ direction, the absorption peak exhibits a monotone increasing tendency during the increase of $n$. At the same time, it can be clearly observed that the increase of the electron doping $n$ of BP also leads to blue shifts of the resonance wavelengths for both $E$ along $x$ and $y$ directions.

\begin{figure}[htbp]
\centering
\includegraphics
[scale=0.35]{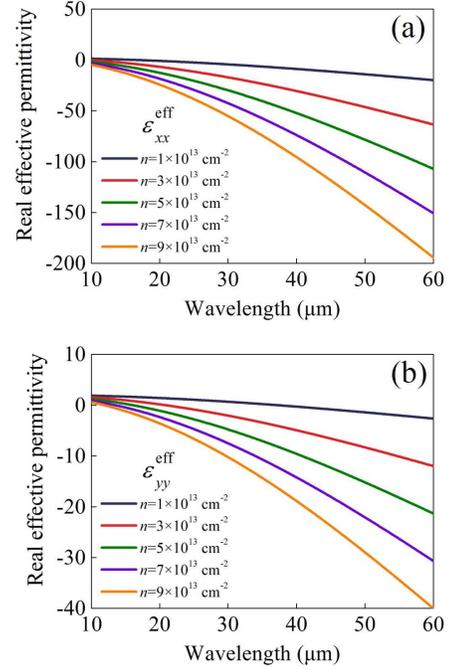}
\caption{\label{fig:7} The real parts of the effective permittivity tensors along (a) $x$ and (b) $y$ directions for the multilayer hyperbolic structure with various electron doping of BP ($n$ ranging from $1\times10^{-13}$ cm$^{-2}$ to $9\times10^{-13}$ cm$^{-2}$).}
\end{figure}

The variation in the absorption peak originates from the change in the effective impedance of the proposed structure influenced by the electron doping of BP. The effective impedances of the whole structure for $E$ along $x$ direction can be calculated as examples. When the electron doping $n$ starts at $1\times10^{13}$ cm$^{-2}$, the effective impedance is $0.12-0.05i$ at 34.57 $\upmu$m, which mismatches to that of the free space and leads to the absorption peak of only 37.78$\%$. As $n$ increases to $5\times10^{13}$ cm$^{-2}$, the effective impedance is $0.99-0.02i$ at 20.41 $\upmu$m, matching to that of the free space, and the perfect absorption is achieved. Finally when $n$ comes to $9\times10^{13}$ cm$^{-2}$, the effective impedance is $2.44-0.38i$ at 17.87 $\upmu$m, and the absorption peak turns down to 81.58$\%$ due to the impedance mismatch. On the other hand, the blue shifts of the resonance wavelengths can be attributed to the fact that the effective permittivities for the multilayer hyperbolic structure decrease as the electron doping of BP increase as shown in Fig.~\ref{fig:7}(a) and (b). Therefore, to approach the value of the effective permittivity $\varepsilon_{xx}^{eff}=-18.87$ and $\varepsilon_{yy}^{eff}=-2.44$ at the perfect absorption, the resonance wavelengths for the cases of $n=1\times10^{13}$ cm$^{-2}$ and $3\times10^{13}$ cm$^{-2}$ need to shift towards longer wavelengths, while the resonance wavelengths for the cases of $n=7\times10^{13}$ cm$^{-2}$ and $9\times10^{13}$ cm$^{-2}$ need to shift towards shorter wavelengths. It is also observed that when the electron doping comes to $n=9\times10^{13}$ cm$^{-2}$, there are bumps at the shorter wavelength side of the absorption peak for $E$ along $x$ direction, which can be attributed to the high order resonances. For the resonance located at 14.31 $\upmu$m, as an example, the relatively weak confinement of electric field between the adjacent unit cells can be observed in the
electric field distribution (not shown). In contrast to the electric dipole resonance, the divergence and convergence of the electric field exhibit a complicated tendency: at the top of the adjacent, the positive and negative charges accumulate at the left and right sides of the gaps, and at the bottom of the adjacent, the positive and negative charges accumulate at the right and left sides of the gaps, showing an opposite distribution, forming the electric quadrupole resonance.  

\begin{figure}[htbp]
\centering
\includegraphics
[scale=0.35]{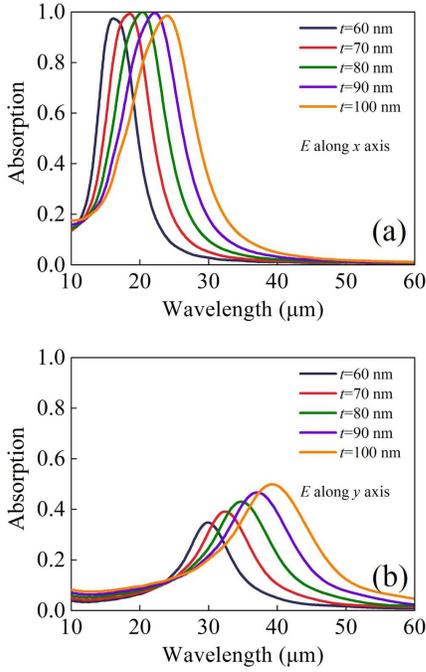}
\caption{\label{fig:8} The anisotropic absorption spectra for electric field $E$ along (a) $x$ and (b) $y$ directions under normal incidence with various thicknesses of the BP/dielectric bilayer ($t$ ranging from 60 nm to 100 nm) in the proposed hyperbolic metamaterials.}
\end{figure}

Next we investigate the absorption responses of the proposed structure with different thickness $t$ of the BP/dielectric bilayer, and the electron doping of BP is fixed as $n=5\times10^{13}$ cm$^{-2}$ and the number of the BP/dielectric bilayer is set as $N=20$. As shown in Fig.~\ref{fig:8}(a), all the absorption peaks of the proposed structure are above 97$\%$ for $E$ along $x$ direction, which indicates that we can approach the near perfect absorption at different wavelengths by adjusting the thickness of the BP/dielectric bilayer. Also, for $E$ along $y$ direction, it can be observed a slight increase in the absorption peak from 34.77$\%$ to 49.88$\%$ as the thickness increases, as Fig.~\ref{fig:8}(b) shows. At the same time, with the increase of the thickness of the BP/dielectric bilayer from $t=60$ nm to 100 nm, the resonance wavelengths for $E$ along both $x$ and $y$ directions show redshifts from 16.10 $\upmu$m to 23.95 $\upmu$m and from 29.95 $\upmu$m to 39.10 $\upmu$m, respectively.

\begin{figure}[htbp]
\centering
\includegraphics
[scale=0.35]{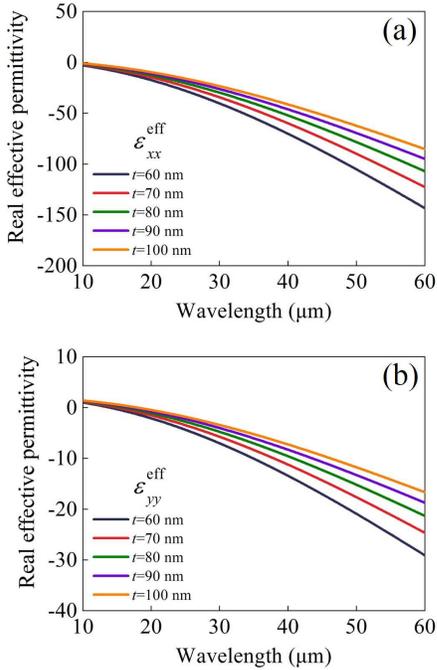}
\caption{\label{fig:9} The real parts of the effective permittivity tensors along (a) $x$ and (b) $y$ directions for the multilayer hyperbolic structure with various thicknesses of the BP/dielectric bilayer ($t$ ranging from 60 nm to 100 nm).}
\end{figure}

The reason for the variation in the absorption peak lies in that the change in the thickness of the BP/dielectric bilayer alters the effective impedance of the proposed structure. For example, the effective impedances of the proposed structure for $E$ along $x$ direction is calculated as $0.72-0.05i$ at 16.10 $\upmu$m for $t=60$ nm, $0.99-0.02i$ at 20.41 $\upmu$m for $t=80$ nm and $1.27-0.02i$ at 23.95 $\upmu$m for $t=100$ nm, respectively. The impedances match to that of free space leads to the nearly perfect absorption of 97.35$\%$, 100$\%$ and 98.51$\%$ at the respective resonance wavelengths. On the other hand, the redshifts of the resonance wavelengths can be explained by the variations of the effective permittivity tensors with the increase of thickness of the BP/dielectric bilayer. In Fig.~\ref{fig:9}(a) and (b), both effective permittivity tensors $\varepsilon_{xx}^{eff}$ along $x$ direction, and $\varepsilon_{yy}^{eff}$ along $y$ direction show the rising tendency for a certain wavelength as the thickness increases from 60 nm to 100 nm. To satisfy the permittivity tensors at the absorption peak, the resonance wavelengths for the cases of $t=60$ nm and 70 nm need to shift towards shorter wavelengths while the resonance wavelengths for the cases of $t=90$ nm and 100 nm need to shift towards longer wavelengths.

\begin{figure}[htbp]
\centering
\includegraphics
[scale=0.35]{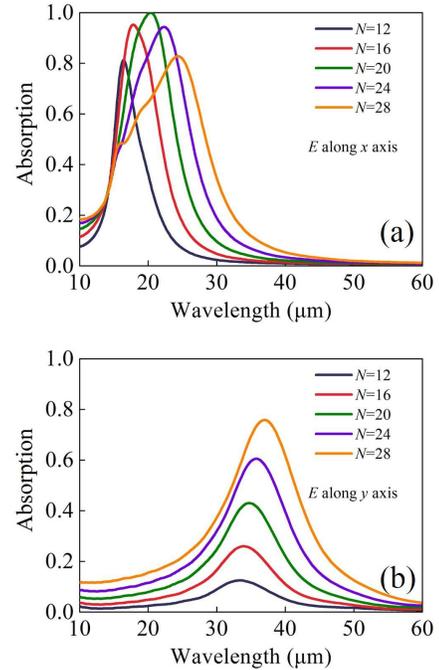}
\caption{\label{fig:10} The anisotropic absorption spectra for electric field $E$ along (a) $x$ and (b) $y$ directions under normal incidence with various numbers of the BP/dielectric bilayer ($N$ ranging from 12 to 28) in the proposed hyperbolic metamaterials.}
\end{figure}

The absorption response of the proposed structure can also be tuned by altering the number of the BP/dielectric bilayer, as shown in Fig.~\ref{fig:10}(a) and (b). For $E$ along $x$ direction, the absorption peak shows an increase from 81.17$\%$ to 100$\%$ when $N$ increases from 12 to 20, and then gradually declines to 82.75$\%$ as $N$ continues increasing to 28. For $E$ along $y$ direction, the absorption peak shows a monotone increase from 12.46$\%$ to 75.82$\%$. These variations can be explained by the reason that the different layer numbers influence the structure of the unit cell and further change the effective impedance of the structure. For example, when $N =12$, 20 and 28, the effective impedances of the whole structure for $E$ along $x$ direction are calculated as $0.40-0.01i$ at resonance wavelength of 16.40 $\upmu$m, $0.99-0.02i$ at 20.41 $\upmu$m, and $2.40-0.16i$ at 24.48 $\upmu$m, respectively. The impedance match of the whole structure to that of free space leads a perfect absorption for $N=20$, while the mismatch causes the relatively weak absorption for $N=12$ and 28. It is also observed that the resonance wavelengths show slight redshifts as the number of the BP/dielectric bilayer increases from $N=12$ to 28 for both $E$ along $x$ and $y$ directions, which can also be attributed to the variations of the effective permittivity tensors. Besides, when the number of BP/dielectric bilayer comes to $N=28$, there are bumps at the shorter wavelength side of the absorption peak for $E$ along $x$ direction, whcih can also be attributed to the high order resonance, as is the case with the high doping level. It needs to be noted that the effective permittivity tensors in Eqs.~(\ref{eq:4})-(\ref{eq:5}) are defined with effective medium theory under the assumption that the number of the BP/dielectric bilayer is infinite. Hence a slight deviation on the effective permittivity tensors would happen when limited layer number is employed to approximate the infinite condition in the practical simulations.

\section{Conclusions}\label{sec4}
In conclusions, we theoretically investigate the tunable anisotropic absorption based on the structure composed of BP/dielectric multilayer stacking unit cells patterned on a gold mirror. The proposed structure shows perfect absorption for $E$ along $x$ direction while the absorption of 8.2$\%$ for $E$ along $y$ direction at the same wavelength. The physical mechanism of the perfect absorption lies in the impedance match associated with the critical coupling condition due to the excitation of the electric dipole resonance between the adjacent unit cells, and the anisotropic responses can be attributed to the asymmetric crystal structure of BP. By changing the electron doping of BP, the thickness and the number of the BP/dielectric bilayer in the unit cell, the absorption responses of the proposed structure can be flexibly controlled. This work demonstrates the potentials of BP as an excellent building block for multilayer structure of hyperbolic metamaterials, and provides inspiration and guidance for a wide variety of tunable anisotropic metadevices such as polarizers and signal processing systems based on hybrid BP/dielectric multilayer structures.


\begin{thebibliography}{10}
	\providecommand{\url}[1]{#1}
	\csname url@samestyle\endcsname
	\providecommand{\newblock}{\relax}
	\providecommand{\bibinfo}[2]{#2}
	\providecommand{\BIBentrySTDinterwordspacing}{\spaceskip=0pt\relax}
	\providecommand{\BIBentryALTinterwordstretchfactor}{4}
	\providecommand{\BIBentryALTinterwordspacing}{\spaceskip=\fontdimen2\font plus
		\BIBentryALTinterwordstretchfactor\fontdimen3\font minus
		\fontdimen4\font\relax}
	\providecommand{\BIBforeignlanguage}[2]{{%
			\expandafter\ifx\csname l@#1\endcsname\relax
			\typeout{** WARNING: IEEEtran.bst: No hyphenation pattern has been}%
			\typeout{** loaded for the language `#1'. Using the pattern for}%
			\typeout{** the default language instead.}%
			\else
			\language=\csname l@#1\endcsname
			\fi
			#2}}
	\providecommand{\BIBdecl}{\relax}
	\BIBdecl
	
	\bibitem{liu2011metamaterials}
	Y.~Liu and X.~Zhang, ``Metamaterials: a new frontier of science and
	technology,'' \emph{Chem. Soc. Rev.}, vol.~40, no.~5, pp. 2494--2507, 2011.
	
	\bibitem{poddubny2013hyperbolic}
	A.~Poddubny, I.~Iorsh, P.~Belov, and Y.~Kivshar, ``Hyperbolic metamaterials,''
	\emph{Nat. Photonics}, vol.~7, no.~12, p. 948, 2013.
	
	\bibitem{sun2014indefinite}
	J.~Sun, N.~M. Litchinitser, and J.~Zhou, ``Indefinite by nature: from
	ultraviolet to terahertz,'' \emph{ACS Photonics}, vol.~1, no.~4, pp.
	293--303, 2014.
	
	\bibitem{jacob2006optical}
	Z.~Jacob, L.~V. Alekseyev, and E.~Narimanov, ``Optical hyperlens: far-field
	imaging beyond the diffraction limit,'' \emph{Opt. Express}, vol.~14, no.~18,
	pp. 8247--8256, 2006.
	
	\bibitem{lu2012hyperlenses}
	D.~Lu and Z.~Liu, ``Hyperlenses and metalenses for far-field super-resolution
	imaging,'' \emph{Nat. Commun.}, vol.~3, p. 1205, 2012.
	
	\bibitem{liang2015squeezing}
	G.~Liang, C.~Wang, Z.~Zhao, Y.~Wang, N.~Yao, P.~Gao, Y.~Luo, G.~Gao, Q.~Zhao,
	and X.~Luo, ``Squeezing bulk plasmon polaritons through hyperbolic
	metamaterials for large area deep subwavelength interference lithography,''
	\emph{Adv. Opt. Mater.}, vol.~3, no.~9, pp. 1248--1256, 2015.
	
	\bibitem{lu2014enhancing}
	D.~Lu, J.~J. Kan, E.~E. Fullerton, and Z.~Liu, ``Enhancing spontaneous emission
	rates of molecules using nanopatterned multilayer hyperbolic metamaterials,''
	\emph{Nat. Nanotechnol.}, vol.~9, no.~1, p.~48, 2014.
	
	\bibitem{roth2017spontaneous}
	D.~J. Roth, A.~V. Krasavin, A.~Wade, W.~Dickson, A.~Murphy, S.~Kéna-Cohen,
	R.~Pollard, G.~A. Wurtz, D.~Richards, S.~A. Maier \emph{et~al.},
	``Spontaneous emission inside a hyperbolic metamaterial waveguide,''
	\emph{ACS Photonics}, vol.~4, no.~10, pp. 2513--2521, 2017.
	
	\bibitem{rustomji2017measurement}
	K.~Rustomji, R.~Abdeddaim, C.~M. de~Sterke, B.~Kuhlmey, and S.~Enoch,
	``Measurement and simulation of the polarization-dependent purcell factor in
	a microwave fishnet metamaterial,'' \emph{Phys. Rev. B}, vol.~95, no.~3, p.
	035156, 2017.
	
	\bibitem{kabashin2009plasmonic}
	A.~Kabashin, P.~Evans, S.~Pastkovsky, W.~Hendren, G.~Wurtz, R.~Atkinson,
	R.~Pollard, V.~Podolskiy, and A.~Zayats, ``Plasmonic nanorod metamaterials
	for biosensing,'' \emph{Nat. Mater.}, vol.~8, no.~11, p. 867, 2009.
	
	\bibitem{sreekanth2016extreme}
	K.~V. Sreekanth, Y.~Alapan, M.~ElKabbash, E.~Ilker, M.~Hinczewski, U.~A.
	Gurkan, A.~De~Luca, and G.~Strangi, ``Extreme sensitivity biosensing platform
	based on hyperbolic metamaterials,'' \emph{Nat. Mater.}, vol.~15, no.~6, p.
	621, 2016.
	
	\bibitem{sreekanth2016enhancing}
	K.~V. Sreekanth, Y.~Alapan, M.~ElKabbash, A.~M. Wen, E.~Ilker, M.~Hinczewski,
	U.~A. Gurkan, N.~F. Steinmetz, and G.~Strangi, ``Enhancing the angular
	sensitivity of plasmonic sensors using hyperbolic metamaterials,'' \emph{Adv.
		Opt. Mater.}, vol.~4, no.~11, pp. 1767--1772, 2016.
	
	\bibitem{zhou2014experiment}
	J.~Zhou, A.~F. Kaplan, L.~Chen, and L.~J. Guo, ``Experiment and theory of the
	broadband absorption by a tapered hyperbolic metamaterial array,'' \emph{ACS
		Photonics}, vol.~1, no.~7, pp. 618--624, 2014.
	
	\bibitem{yin2015ultra}
	X.~Yin, L.~Chen, and X.~Li, ``Ultra-broadband super light absorber based on
	multi-sized tapered hyperbolic metamaterial waveguide arrays,'' \emph{J.
		Lightwave Technol.}, vol.~33, no.~17, pp. 3704--3710, 2015.
	
	\bibitem{chang2016metasurface}
	Y.-C. Chang, A.~V. Kildishev, E.~E. Narimanov, and T.~B. Norris, ``Metasurface
	perfect absorber based on guided resonance of a photonic hypercrystal,''
	\emph{Phys. Rev. B}, vol.~94, no.~15, p. 155430, 2016.
	
	\bibitem{xia2014two}
	F.~Xia, H.~Wang, D.~Xiao, M.~Dubey, and A.~Ramasubramaniam, ``Two-dimensional
	material nanophotonics,'' \emph{Nat. Photonics}, vol.~8, no.~12, p. 899,
	2014.
	
	\bibitem{he2016further}
	X.~He, P.~Gao, and W.~Shi, ``A further comparison of graphene and thin metal
	layers for plasmonics,'' \emph{Nanoscale}, vol.~8, no.~19, pp.
	10\,388--10\,397, 2016.
	
	\bibitem{xiao2017strong}
	S.~Xiao, T.~Wang, X.~Jiang, X.~Yan, L.~Cheng, B.~Wang, and C.~Xu, ``Strong
	interaction between graphene layer and fano resonance in terahertz
	metamaterials,'' \emph{J. Phys. D: Appl. Phys.}, vol.~50, no.~19, p. 195101,
	2017.
	
	\bibitem{xiao2018active}
	S.~Xiao, T.~Wang, T.~Liu, X.~Yan, Z.~Li, and C.~Xu, ``Active modulation of
	electromagnetically induced transparency analogue in terahertz hybrid
	metal-graphene metamaterials,'' \emph{Carbon}, vol. 126, pp. 271--278, 2018.
	
	\bibitem{li2018wavelength}
	H.-J. Li, Y.-Z. Ren, J.~Hu, M.~Qin, and L.~Wang, ``Wavelength-selective
	wide-angle light absorption enhancement in monolayers of transition-metal
	dichalcogenides,'' \emph{J. Lightwave Technol.}, vol.~36, no.~16, pp.
	3236--3241, 2018.
	
	\bibitem{iorsh2013hyperbolic}
	I.~V. Iorsh, I.~S. Mukhin, I.~V. Shadrivov, P.~A. Belov, and Y.~S. Kivshar,
	``Hyperbolic metamaterials based on multilayer graphene structures,''
	\emph{Phys. Rev. B}, vol.~87, no.~7, p. 075416, 2013.
	
	\bibitem{chang2016realization}
	Y.-C. Chang, C.-H. Liu, C.-H. Liu, S.~Zhang, S.~R. Marder, E.~E. Narimanov,
	Z.~Zhong, and T.~B. Norris, ``Realization of mid-infrared graphene hyperbolic
	metamaterials,'' \emph{Nat. Commun.}, vol.~7, p. 10568, 2016.
	
	\bibitem{sreekanth2013negative}
	K.~Sreekanth, A.~De~Luca, and G.~Strangi, ``Negative refraction in
	graphene-based hyperbolic metamaterials,'' \emph{Appl. Phys. Lett.}, vol.
	103, no.~2, p. 023107, 2013.
	
	\bibitem{barzegar2016study}
	S.~Barzegar-Parizi, ``Study of backward waves in multilayered structures
	composed of graphene micro-ribbons,'' \emph{J. Appl. Phys.}, vol. 119,
	no.~19, p. 193105, 2016.
	
	\bibitem{othman2013graphene}
	M.~A. Othman, C.~Guclu, and F.~Capolino, ``Graphene-based tunable hyperbolic
	metamaterials and enhanced near-field absorption,'' \emph{Opt. Express},
	vol.~21, no.~6, pp. 7614--7632, 2013.
	
	\bibitem{he2013broadband}
	S.~He and T.~Chen, ``Broadband thz absorbers with graphene-based anisotropic
	metamaterial films,'' \emph{IEEE Trans. THz Sci. Technol.}, vol.~3, no.~6,
	pp. 757--763, 2013.
	
	\bibitem{su2015terahertz}
	Z.~Su, J.~Yin, and X.~Zhao, ``Terahertz dual-band metamaterial absorber based
	on graphene/mgf 2 multilayer structures,'' \emph{Opt. Express}, vol.~23,
	no.~2, pp. 1679--1690, 2015.
	
	\bibitem{jia2015tunable}
	Y.~Jia, H.~Zhao, Q.~Guo, X.~Wang, H.~Wang, and F.~Xia, ``Tunable
	plasmon--phonon polaritons in layered graphene--hexagonal boron nitride
	heterostructures,'' \emph{ACS Photonics}, vol.~2, no.~7, pp. 907--912, 2015.
	
	\bibitem{tyszka2017tunable}
	A.~Tyszka-Zawadzka, B.~Janaszek, and P.~Szczepa{\'n}ski, ``Tunable slow light
	in graphene-based hyperbolic metamaterial waveguide operating in sclu telecom
	bands,'' \emph{Opt. Express}, vol.~25, no.~7, pp. 7263--7272, 2017.
	
	\bibitem{low2014tunable}
	T.~Low, A.~Rodin, A.~Carvalho, Y.~Jiang, H.~Wang, F.~Xia, and A.~C. Neto,
	``Tunable optical properties of multilayer black phosphorus thin films,''
	\emph{Phys. Rev. B}, vol.~90, no.~7, p. 075434, 2014.
	
	\bibitem{ling2015renaissance}
	X.~Ling, H.~Wang, S.~Huang, F.~Xia, and M.~S. Dresselhaus, ``The renaissance of
	black phosphorus,'' \emph{Proc. Nat. Acad. Sci. USA}, vol. 112, no.~15, pp.
	4523--4530, 2015.
	
	\bibitem{liu2018dynamical}
	X.~Liu, W.~Lu, X.~Zhou, Y.~Zhou, C.~Zhang, J.~Lai, S.~Ge, M.~C. Sekhar, S.~Jia,
	K.~Chang \emph{et~al.}, ``Dynamical anisotropic response of black phosphorus
	under magnetic field,'' \emph{2D Mater.}, vol.~5, no.~2, p. 025010, 2018.
	
	\bibitem{low2014plasmons}
	T.~Low, R.~Rold{\'a}n, H.~Wang, F.~Xia, P.~Avouris, L.~M. Moreno, and
	F.~Guinea, ``Plasmons and screening in monolayer and multilayer black
	phosphorus,'' \emph{Phys. Rev. Lett.}, vol. 113, no.~10, p. 106802, 2014.
	
	\bibitem{gonccalves2017hybridized}
	P.~A.~D. Gon{\c{c}}alves, S.~Xiao, N.~Peres, and N.~A. Mortensen, ``Hybridized
	plasmons in 2d nanoslits: From graphene to anisotropic 2d materials,''
	\emph{ACS Photonics}, vol.~4, no.~12, pp. 3045--3054, 2017.
	
	\bibitem{lu2017strong}
	H.~Lu, Y.~Gong, D.~Mao, X.~Gan, and J.~Zhao, ``Strong plasmonic confinement and
	optical force in phosphorene pairs,'' \emph{Opt. Express}, vol.~25, no.~5,
	pp. 5255--5263, 2017.
	
	\bibitem{nong2018strong}
	J.~Nong, W.~Wei, W.~Wang, G.~Lan, Z.~Shang, J.~Yi, and L.~Tang, ``Strong
	coherent coupling between graphene surface plasmons and anisotropic black
	phosphorus localized surface plasmons,'' \emph{Opt. Express}, vol.~26, no.~2,
	pp. 1633--1644, 2018.
	
	\bibitem{qing2018tailoring}
	Y.-M. Qing, H.-F. Ma, and T.-J. Cui, ``Tailoring anisotropic perfect absorption
	in monolayer black phosphorus by critical coupling at terahertz
	frequencies,'' \emph{Opt. Express}, vol.~26, no.~25, pp. 32\,442--32\,450,
	2018.
	
	\bibitem{lu2015bandgap}
	J.~Lu, J.~Wu, A.~Carvalho, A.~Ziletti, H.~Liu, J.~Tan, Y.~Chen, A.~Castro~Neto,
	B.~Ozyilmaz, and C.~H. Sow, ``Bandgap engineering of phosphorene by laser
	oxidation toward functional 2d materials,'' \emph{ACS Nano}, vol.~9, no.~10,
	pp. 10\,411--10\,421, 2015.
	
	\bibitem{correas2016black}
	D.~Correas-Serrano, J.~Gomez-Diaz, A.~A. Melcon, and A.~Al{\`u}, ``Black
	phosphorus plasmonics: anisotropic elliptical propagation and
	nonlocality-induced canalization,'' \emph{J. Opt.}, vol.~18, no.~10, p.
	104006, 2016.
	
	\bibitem{li2017tunable}
	D.~Li, J.-R. Xu, K.~Ba, N.~Xuan, M.~Chen, Z.~Sun, Y.-Z. Zhang, and Z.~Zhang,
	``Tunable bandgap in few-layer black phosphorus by electrical field,''
	\emph{2D Mater.}, vol.~4, no.~3, p. 031009, 2017.
	
	\bibitem{liu2016localized}
	Z.~Liu and K.~Aydin, ``Localized surface plasmons in nanostructured monolayer
	black phosphorus,'' \emph{Nano Lett.}, vol.~16, no.~6, pp. 3457--3462, 2016.
	
	\bibitem{xiong2017strong}
	F.~Xiong, J.~Zhang, Z.~Zhu, X.~Yuan, and S.~Qin, ``Strong anisotropic perfect
	absorption in monolayer black phosphorous and its application as tunable
	polarizer,'' \emph{J. Opt.}, vol.~19, no.~7, p. 075002, 2017.
	
	\bibitem{song2018biaxial}
	X.~Song, Z.~Liu, Y.~Xiang, and K.~Aydin, ``Biaxial hyperbolic metamaterials
	using anisotropic few-layer black phosphorus,'' \emph{Opt. Express}, vol.~26,
	no.~5, pp. 5469--5477, 2018.
	
	\bibitem{yuan2018highly}
	Y.~Yuan, X.~Yu, Q.~Ouyang, Y.~Shao, J.~Song, J.~Qu, and K.-T. Yong, ``Highly
	anisotropic black phosphorous-graphene hybrid architecture for
	ultrassensitive plasmonic biosensing: Theoretical insight,'' \emph{2D
		Mater.}, vol.~5, no.~2, p. 025015, 2018.
	
	\bibitem{hong2018towards}
	Q.~Hong, F.~Xiong, W.~Xu, Z.~Zhu, K.~Liu, X.~Yuan, J.~Zhang, and S.~Qin,
	``Towards high performance hybrid two-dimensional material plasmonic devices:
	strong and highly anisotropic plasmonic resonances in nanostructured
	graphene-black phosphorus bilayer,'' \emph{Opt. Express}, vol.~26, no.~17,
	pp. 22\,528--22\,535, 2018.
	
	\bibitem{feng2019perfect}
	N.~Feng, J.~Zhu, C.~Li, Y.~Zhang, Z.~Wang, Z.~Liang, and Q.~H. Liu, ``Perfect
	anisotropic infrared absorption in monolayer black phosphorus with/without
	subwavelength patterning,'' \emph{IEEE J. Sel. Topics Quantum Electron.},
	vol.~25, no.~3, p. 4700407, 2019.
	
	\bibitem{ordal1985optical}
	M.~A. Ordal, R.~J. Bell, R.~W. Alexander, L.~L. Long, and M.~R. Querry,
	``Optical properties of fourteen metals in the infrared and far infrared: Al,
	co, cu, au, fe, pb, mo, ni, pd, pt, ag, ti, v, and w.'' \emph{Appl. Opt.},
	vol.~24, no.~24, pp. 4493--4499, 1985.
	
	\bibitem{agranovich1985notes}
	V.~Agranovich and V.~Kravtsov, ``Notes on crystal optics of superlattices,''
	\emph{Solid State Commun.}, vol.~55, no.~1, pp. 85--90, 1985.
	
	\bibitem{piper2014total}
	J.~R. Piper and S.~Fan, ``Total absorption in a graphene monolayer in the
	optical regime by critical coupling with a photonic crystal guided
	resonance,'' \emph{ACS Photonics}, vol.~1, no.~4, pp. 347--353, 2014.
	
	\bibitem{jiang2017tunable}
	X.~Jiang, T.~Wang, S.~Xiao, X.~Yan, and L.~Cheng, ``Tunable
	ultra-high-efficiency light absorption of monolayer graphene using critical
	coupling with guided resonance,'' \emph{Opt. Express}, vol.~25, no.~22, pp.
	27\,028--27\,036, 2017.
	
	\bibitem{smith2005electromagnetic}
	D.~Smith, D.~Vier, T.~Koschny, and C.~Soukoulis, ``Electromagnetic parameter
	retrieval from inhomogeneous metamaterials,'' \emph{Phys. Rev. E.}, vol.~71,
	no.~3, p. 036617, 2005.
	
	\bibitem{szabo2010unique}
	Z.~Szab{\'o}, G.-H. Park, R.~Hedge, and E.-P. Li, ``A unique extraction of
	metamaterial parameters based on kramers--kronig relationship,'' \emph{IEEE
		Trans. Microw. Theory Techn.}, vol.~58, no.~10, pp. 2646--2653, 2010.
	
\end{thebibliography}


%

%
%
%




\end{document}